# Trapping and Electrical Characterization of Single Core/Shell Iron-Based Nanoparticles in Self-Aligned Nanogaps.

Jacqueline Labra-Muñoz,[1,2] Zorica Konstantinovic,[3] Lluis Balcells,[4] Alberto Pomar,[4] Herre S. J. van der Zant,[1] and Diana Dulić[5, 1]

[1)] *Kavli Institute of Nanoscience, Delft University of Technology, Lorentzweg 1, 2628 CJ Delft, The Netherlands.*

[2)] *Electrical Engineering Department, Faculty of Physical and Mathematical Sciences, University of Chile, Av. Tupper 2007, Santiago, Chile.*

[3)] *Center for Solid State Physics and New Materials, Institute of Physics Belgrade, University of Belgrade, Pregrevica 118, 11080 Belgrade, Serbia.*

[4)] *Institut de Ciència de Materials de Barcelona, ICMAB-CSIC, Campus de la UAB, 08193 Bellaterra, Spain.*

[5)] *Physics Department, Faculty of Physical and Mathematical Sciences, University of Chile, Av. Blanco Encalada 2800, Santiago, Chile.*

We report on the fabrication and measurements of platinum-self-aligned nanogap devices containing cubed iron (core) / iron oxide (shell) nanoparticles (NPs) with two averaged different sizes (13 and 17 nm). The nanoparticles are deposited by means of a cluster gun technique. Their trapping across the nanogap is demonstrated by comparing the current vs. voltage characteristics (*I-V*s) before and after the deposition. At low temperature, the *I-V*s can be well fitted to the Korotkov and Nazarov Coulomb blockade model, which captures the coexistence of single-electron tunneling and tunnel barrier suppression upon a bias voltage increase. The measurements thus show that Coulomb-blockaded devices can be made with a nanoparticle cluster source, which extends the existing possibilities to fabricate such devices to those in which it is very challenging to reduce the usual NPs agglomeration given by a solution method.

Due to the development in fabrication techniques in the last decades, it is now possible to realize nano-electronic devices with an electrode spacing down to the nanometer scale. In combination with their optical and magnetic properties, the unique size-dependent charge transport properties of nanoparticles (NPs) make them interesting candidates for exploring functionalities in such devices including those associated with biomedical applications[1–4]. In this respect, iron oxide NPs represent intriguing examples. From a magnetic perspective, magnetite ($Fe_3O_4$) exhibits the strongest magnetism of any transition metal oxide[5]. At room temperature, bulk magnetite is ferrimagnetic. However, at the same temperature, magnetite particles of a few nanometers in size are superparamagnetic. This aspect makes magnetite NPs suitable for use in magnetic resonance imaging (MRI) contrast agents for molecular and cell imaging[5,6]. In addition, self-assembled iron-oxide NPs are proposed as data storage devices[7,8], being potential key components for a new generation of electronic materials[9,10].

Electrical characterization of NPs on a single-particle level implies two major challenges: (i) the fabrication of electrodes with a separation (gap) of a few nanometers, so that single NPs bridge the gap from source to drain and (ii) the synthesis and deposition of reproducible NPs (in size and density) in the nanogaps. To decide which nano-electrode fabrication technique to use depends on the NP shape, size, composition and the specific research aim. Thus far, the methods for trapping of NPs in nanogaps involve the deposition from a solution[11–16], and among them the drop-casting technique is the most common[12,13]. It can be used in combination with a subsequent drying process, such as exposure to high temperatures[14], or vacuum exposure[15] or in combination with applying an electric field (electrophoresis)[16]. The advantage of drop-casting is that it represents a very simple method[17]; however, the usual NP agglomeration by the drop casting method can make a controlled deposition on the surface of the device challenging[18].

In this work, we studied core/shell $Fe/Fe_3O_4$ nanoparticles, that are deposited on self-aligned nanogaps by means of a non-solution based, cluster source[19]. The method offers excellent control of the size distribution and stoichiometry of the NPs while minimizing NP agglomeration[20]. This constitutes the realization of devices in which single NPs are contacted in nanogaps using this deposition technique, which has not been reported before. We find that the devices are stable and allow for electrical characterization at room and low temperatures showing Coulomb blockade coexisting with barrier suppression as the main transport mechanism.

A schematic of the nanogap chip design is shown in Fig. 1. It consists of 36 devices, formed by a main electrode (in yellow) and 36 finger-like-auxiliary electrodes (in gray); each finger-like electrode has a length of 5 $\mu$m and width of 1 $\mu$m. The gap between the main electrode and each auxiliary electrode (device) varies between 12 and 21 nm [see Fig. 1(b)]. The devices are enumerated from 1 to 36, as illustrated in Figure 1(a). The self-aligned nanogaps are not defined by direct e-beam writing, but instead are the result of a mask formed by chromium oxidation[21–23] (see end of the document for details). The nanoparticles have a cubic shape and consist of an iron core covered with an iron oxide shell ($Fe_3O_4$)[24]; see Supplementary Material (SM), Fig. S5. Specifically, we measured two chips with NPs that differ in size; the average sizes of the NPs are 13 nm (denoted chip *Small NPs*) and 17 nm (denoted chip *Big NPs*), respectively. Fig. 1(d) shows a transmission electron microscopy (TEM) image of *Big NPs* from the same batch as used for the deposition. The particles are synthesized by a cluster source and in-situ deposited on the devices with previously patterned electrode structures. After deposition, the samples are taken out of the chamber and placed in a probe station for further electrical characterization.

[1] Electronic mail: ddulic@ing.uchile.cl

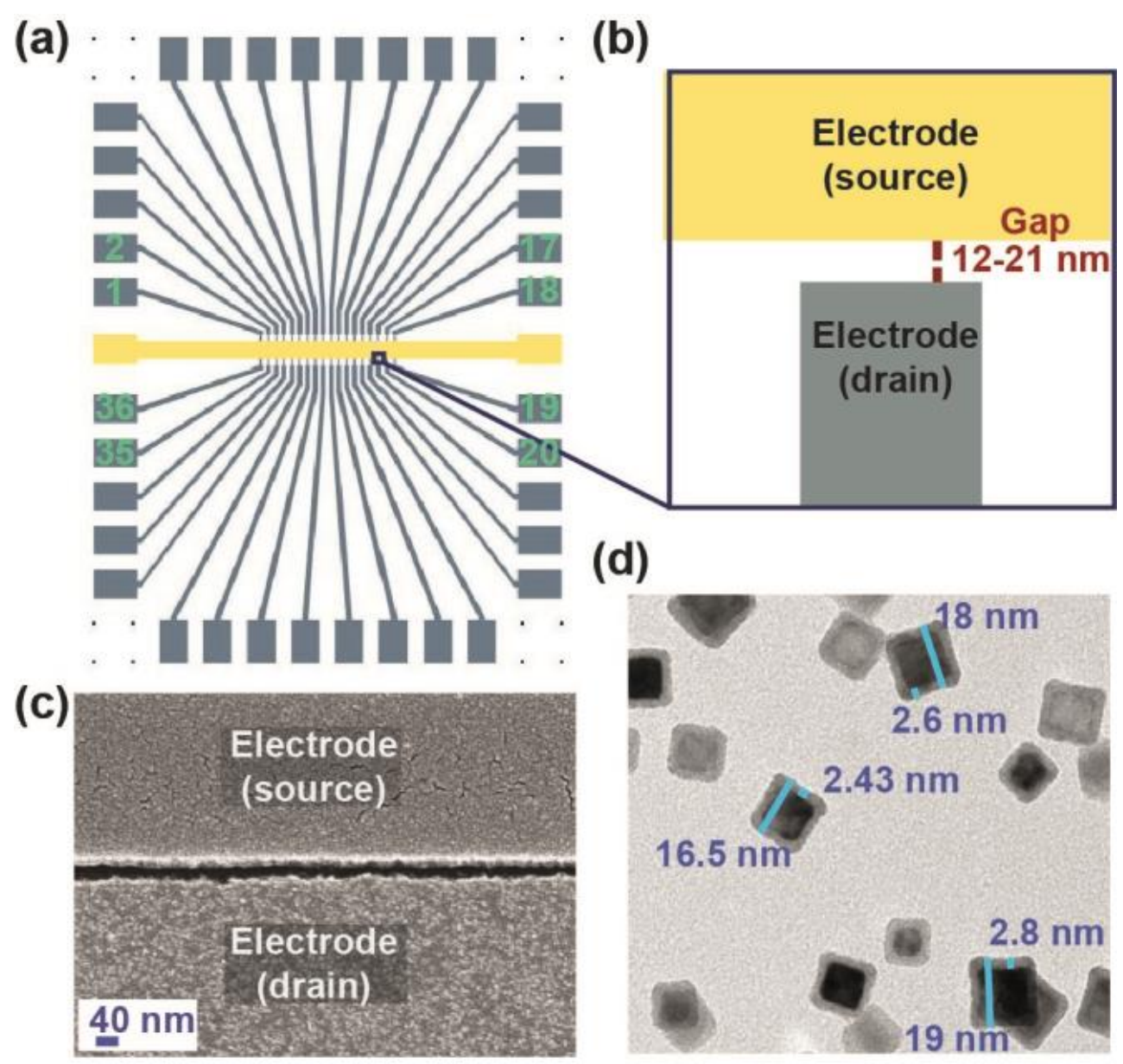

FIG. 1. (a) General design of the chip. In yellow, the main electrode is represented as the source. In grey, 36 auxiliary electrodes are shown, represented as the drain. (b) Schematic of the gap between a pair of source and drain electrodes (device). (c) Scanning electron microscopy image of an empty device. (d) Transmission electron microscopy image of the iron (core) / iron oxide (shell) nanoparticles (*Big NPs*) from the same batch as used for the deposition.

Prior to NP deposition, the current vs. voltage (*I-V*) characteristic of each electrode pair was recorded [Fig. 2(a)]. The noise level in our probe-station measurements was about 1 pA. We have chosen twice this value (i.e., 2 pA) as the threshold value to determine if NP trapping occurred in the gap. Thus, a device exhibiting an increase in current greater than 2 pA over the bias voltage range probed (± 1.5 V) was discarded, i.e., only open gaps (called *working devices*) were selected to characterize the NP device (100% of total electrode pairs of the chip *Big NPs*, and 97% of the chip *Small NPs*). Once the NPs were deposited, we identified their presence within the gap [Fig. 2(b)] by comparing the *I-V* curve of the gap before and after deposition, measured in air and at room temperature. Fig. 2(c) shows a typical *I-V* curve measured for device #6 (chip *Big NPs*), with the same appearance as the one presented in Fig. 1(c). After deposition, 92% of the *working devices* on the chip *Big NPs* showed an increase in the current without being short-circuited [Fig. S3(b)], indicating the trapping of NPs between the electrodes. Note that, the *I-V*s show a superlinear behavior at high bias voltage; the current increases faster than the bias voltage does. The percentage of *working devices* on the chip *Small NPs* that trapped NPs after the deposition, was 100% [Fig. S3(a)].

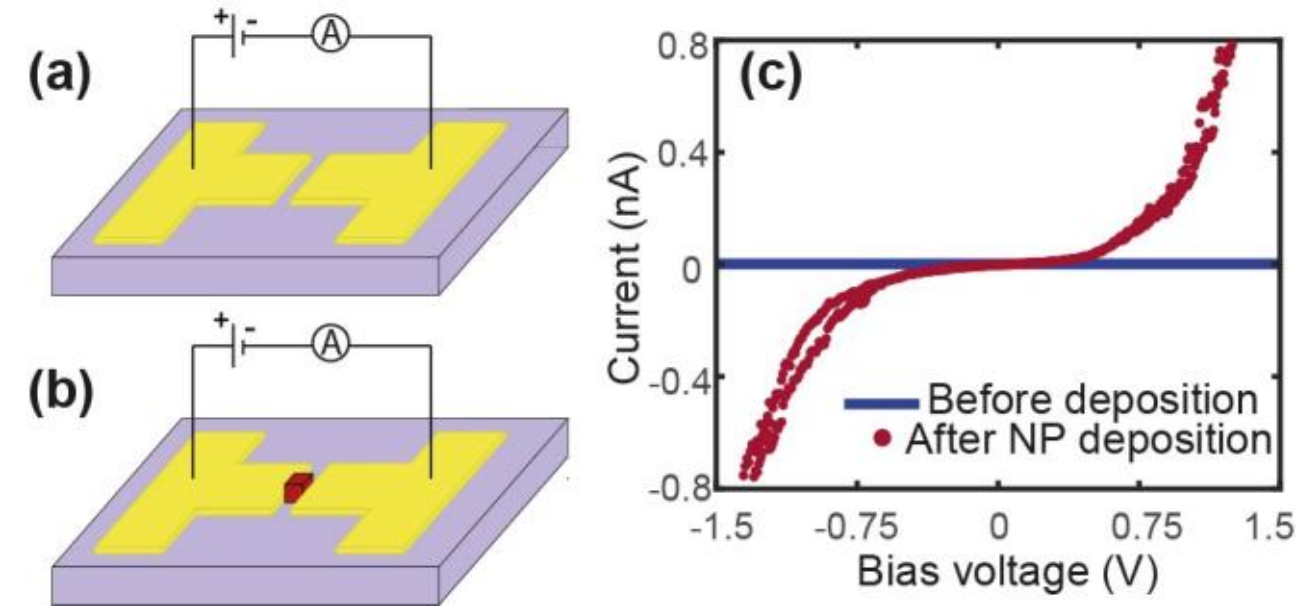

FIG. 2. Description of the measurement procedure. (a) Schematic circuit of a device before nanoparticle (NP) deposition (empty gap). (b) Schematic circuit of a device with an iron NP trapped between the electrodes. (c) Electrical characterization of device #6 (*Big NPs*) before and after NP deposition, measured at room temperature, in vacuum. The blue curve describes an open circuit, reflecting an empty device. The increase in current shown in the red curve indicates the capture of iron NPs. In both cases the current is measured as follows: (I) Voltage sweep from 0 V to 1.5 V. (II) Voltage sweep from 1.5 V to -1.5 V. (III) Voltage sweep from -1.5 V to 0 V.

The NP *working devices* were stable to allow measurements at low temperature (20 K). At this temperature, 40% of the devices on the chip *Big NPs* showed symmetric *I-V*s and 58% of the devices showed asymmetric *I-V*s. For 2% of the devices, the current dropped below the noise level (2 pA) at this temperature over the bias voltage range probed (-1.5 V to 1.5 V). In the case of chip *Small NPs*, only 11% of the devices had symmetric *I-V*s, 49% showed asymmetric *I-V*s, and 40% of the devices showed currents below the threshold value of 2pA. Figure 3 displays four typical symmetric *I-V* curves (in light blue) measured at 20 K, in vacuum, (#2 and #36 of chip *Big NPs*, and #17 and #25 of chip *Small NPs*). For clarity, these *I-V* curves are the descendent curves of the *I-V* cycles, i.e., the current recorded from 1.5 V to −1.5 V. The *I-V*s were found to be free of hysteresis. The observed asymmetry in the other devices (see SM, Fig. S4) may result from an asymmetry in the contact configuration on either side of the junction.

Since the gap and nanoparticles are of the same size (12-21 nm) and the electrode width is 1 $\mu$m, the presence of more than one NP connected in parallel is plausible, although the dominant conductance path way may well be through one particle connected with the lowest tunnel barriers to the two electrodes. With this picture in mind, we used the Korotkov and Nazarov (K-N)[25] model to describe the *I-V* characteristics. This model treats the coexistence of single-electron tunneling and effective tunnel barrier suppression (when increasing the voltage). Bezryadin *et al.*[26] applied this model to describe transport through palladium nanocrystals connected in between electrodes by electrostatic trapping.

According to the K-N model, the tunneling rates expressed in terms of the current at a given temperature T are approximated by the Stratton formula:[27]

$$I(V) = (2\pi k_B T/eR_0)[sinh(eV\tau/\hbar)/sin(2\pi\tau k_B T/\hbar)], \qquad (1)$$

where $\tau = L/\sqrt{2U/m}$ is the tunneling transversal time. $L$ and $U$ are the barrier width and height, respectively. $R_0$ is the resistance of the junction at zero bias and zero temperature, $\hbar$ is the Planck's constant, and $m$ is the electron mass. Unlike the classic Coulomb Blockade model[28], the K-N model captures an essential part of the data, namely the curvature of the *I-V* at higher bias, which is represented by the fitting parameter $\alpha = E_C \cdot \tau/\hbar$, defined as the ratio between the charging energy ($E_C$) and the energy scale for which the barrier suppression takes place. The charging energy is defined as $E_C = e^2/2C$, where $C$ is the total capacitance. To limit the number of fit parameters, we assumed (i) the residual charge induced on the NP to be zero and (ii) the capacitances and resistance on the right and left sides to be equal ($C_1 = C_2, R_1 = R_2$), i.e., the condition for fitting symmetric *I-V* characteristics. Thus, the fitting parameters are $\alpha$, $V_C = e/C$ and $R_0 = \tilde{R}\exp(2L\sqrt{2mU}/\hbar)$, where for *Big NPs*, $\tilde{R}$ is approximated to be the ratio between the quantum resistance (13 kΩ) and the number of quantum channels, which is ~10 considering the NP size.

The symmetric *I-V*s fitted to this model were from 14 *Big NPs*, and 4 *Small NPs* devices. The dark blue curves in Fig. 3 are the K-N fits to the data. The fitting parameters of all symmetric fitted curves are listed in Table S2. The average of the parameter $\alpha$ is 0.54 and 0.62 for *Big* and for *Small NPs* respectively, consistent with the presence of barrier suppression and the associated exponential-like shape of the *I-V* curves. The average values for $V_C$ and $R_0$ are 0.15 V and 3.1 MΩ for the *Big NPs*, while they are 0.22 V and 40.3 MΩ for the *Small NPs*. From these fitting parameters, the height and the width of the tunnel barriers can be estimated, according to the expressions $U = eV_C\, ln(R_0/\tilde{R})/8\alpha$ and $L = \hbar\sqrt{\alpha\, ln(R_0/\tilde{R})/emV_c)}$. The average value of $U$ for the *Big NPs* and *Small NPs* is then found to be 0.3 eV assuming $\tilde{R}$ to be 47 $k$Ω for *Small NPs*, and the average of the estimated $L$ for *Big NPs* and *Small NPs* is 1.5 nm and 1.2 nm, respectively. It can be noticed that $L$ is of the same order of magnitude as the thickness of the iron-oxide shell.

Additionally, from the fits of *Big NPs*, the average total capacitance $C = e/V_C$ is found to be 1.1 aF with a corresponding charging energy of 75 meV. On the other hand, the fits of *Small NPs* devices yield an average $C$ of 0.7 aF and a charging energy of 110 meV, corroborating the fact that the capacitances scales with the particle size. Furthermore, we can compare the estimated capacitances to an upper and a lower bound estimates of the NP capacitance using two parallel plate capacitors located between the iron core of the NP and the two electrodes on either side, connected in series (see supplementary material, Figure S1). One can express those capacitances as $C_{shell1} = C_{shell2} = \varepsilon_r\varepsilon_0 A/d$, where $\varepsilon_0$ is the vacuum permittivity, $\varepsilon_r$ the relative permittivity of the $Fe_3O_4$ shell, which according to Hotta *et al.*[29] can be estimated to be around 8, $d$ is the distance between the plates, which corresponds to the iron-oxide-shell thickness (2.4 nm). The upper bound estimate considers the contact area to be maximized, i.e., the area of the parallel plate *A* is estimated to be 17 x 17 nm$^2$ for *Big NPs* and 13 x 13 nm$^2$ for *Small NPs*. Thus, the estimated capacitance of the nanoparticle is given by $C_{est} = ({C_{shell1}}^{-1} + {C_{shell2}}^{-1})^{-1}$, which results in 4.3 aF for *Big NPs* and 2.5 aF for *Small NPs*. Following an analogous reasoning, the lower limit case considers a minimized contact area (*A*) estimated to be 17 x 2.6 nm$^2$ for *Big NPs* and 13 x 2.6 nm$^2$ for *Small NPs*. The corresponding capacitances are 0.7 aF and 0.5 aF for *Big* and *Small NPs*, respectively. The capacitance obtained from the K-N model lies in between the two estimated limiting values. See Supplementary Material, section I for a more elaborate discussion on the capacitances. Although the number of NPs present in the gaps cannot be established, the consistency between the measurements and the K-N model suggests that the dominant conduction pathway is through one particle. In some cases, like Fig. S2 device #17 (*Small NPs*), SEM images provide an additional indication for this. However, it was not possible to image all measured devices. In case that more particles would contribute, the estimates for the capacitance would not be affected, provided that the offset charge is similar for all of them.

In conclusion, we have demonstrated that individual NPs can be trapped in self-aligned nanogaps using a cluster gun technique to deposit the NPs. The NP devices are stable at low and room temperatures. Electrical characterization shows the *I-V* curves that are consistent with single electron tunneling in combination with barrier suppression to account for the exponential-like shape observed at high bias. The fabrication method can be extended to the study of other types of NPs with the advantage that the direct deposition in vacuum conditions circumvents agglomeration of particles.

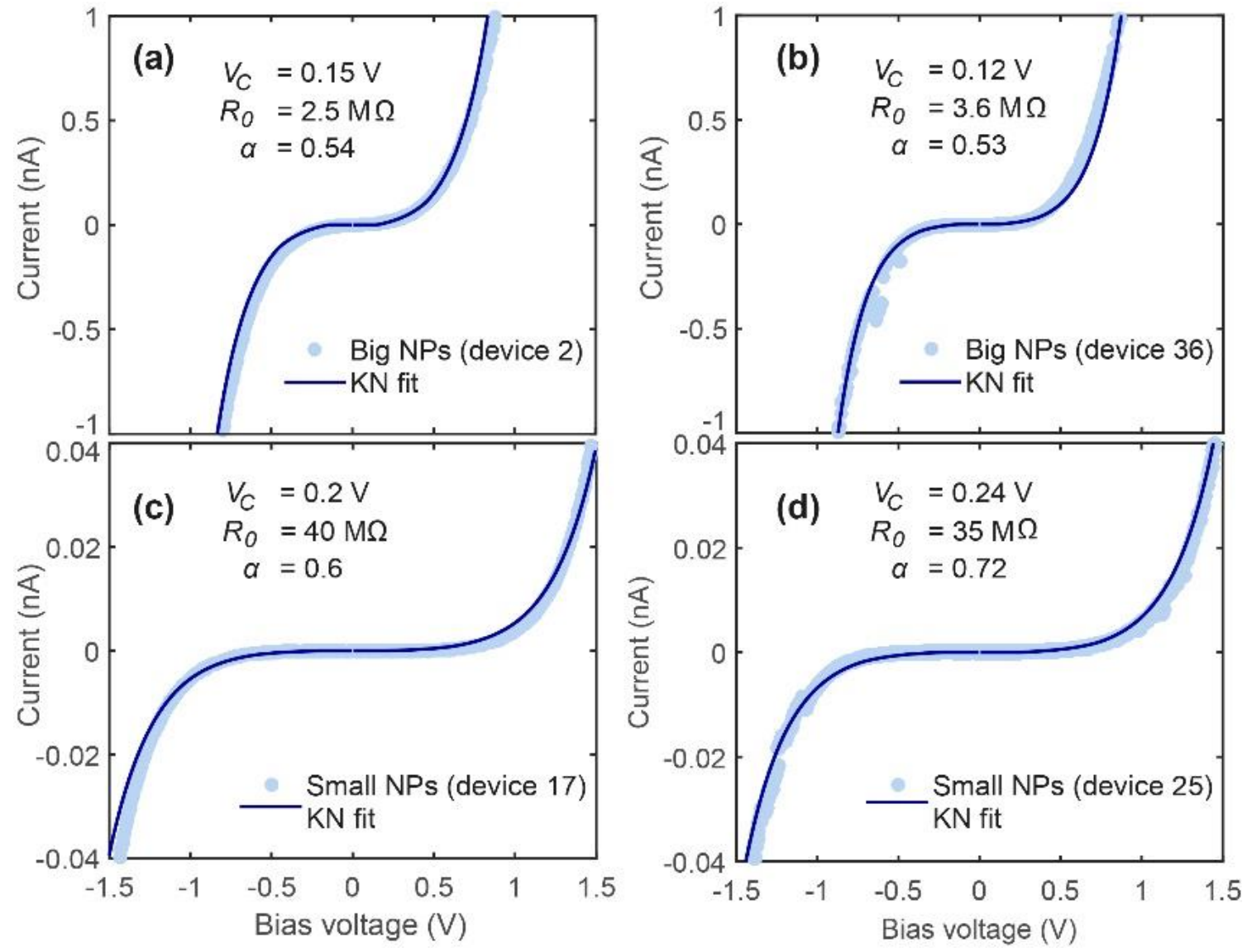

FIG. 3. Symmetric *I-V* characteristics (descendent part (defined in caption Fig. 2c) of the cycle) measured at 20 K, in vacuum. (a) Devices #2 and (b) #36 contain *Big NPs*. (c) Devices #17 and (d) # 25 *Small NPs*. Fit parameters are listed in the inset. The associated charging energies ($E_C$) are 75 meV, 60 meV, 100 meV, and 120 meV, respectively.

The devices are fabricated as follows. On top of a $Si/SiO_2$ substrate, the main electrode is defined by e-beam lithography (EBL), and evaporation of 5 nm of titanium (adhesive layer) and subsequently 30 nm of platinum. On top of the platinum layer, a 25 nm chromium layer is deposited. Upon exposure to ambient conditions, the chromium layer naturally oxidizes, expanding its size. In this manner, chromium oxide acts as a shadow mask of a few nanometers near the edge of the main electrode. The thickness of the chromium layer determines the size of the gap. A second EBL cycle defines the finger-like-auxiliary electrodes, by depositing 5 nm of titanium and 20 nm of platinum. In the final step the chromium layer is etched away (wet-etch step) to reveal the underlying nanogaps. The recipe is depicted in Figure S9.

The NPs are synthesized and deposited by means of a home-built combination of a magnetron sputtering and gas aggregation techniques[19]. A DC magnetron with an Fe target (99.95% purity) was operated typically at 30 W. Deposition took place at a nozzle-substrate distance of 15 cm with a constant Ar flux of 90 sccm and pressures in the low $10^{-3}$ Torr range. To characterize the NPs (particle size and structure), test substrates are placed next to the chip. Si wafers were used for SEM inspection, and carbon-coated grids were used for TEM inspection. The characterization of devices was realized by scanning electron microscopy (SEM) using a QUANTA FEI 200 FEGESEM microscope. The core-shell structure of $Fe/Fe_3O_4$ nanoparticles (crystallinity, morphology, size) was examined by transmission electron microscopy (TEM) using a JEOL, JEM 1210 transmission electron microscope operating at 120 kV. Diffraction patterns of power spectra were obtained from selected regions in the micrographs.

The electrical measurements were performed in a vacuum flow cryostat probe station with TU Delft home-built low-noise electronics. The minimum temperature is around 10-20 K.

See the supplementary material (SM) for more details of this study regarding device fabrication, nanoparticle deposition, and additional results.

This study was supported by the EU Horizon 2020 research and innovation program under the Marie-Sklodowska-Curie Grant Agreement No 645658 (DAFNEOX Project), by two FONDECYT REGULAR grants 1181080 and 1161775, and by two FONDEQUIP grants EQM140055 and EQM180009. We thank Spanish Ministry of Science, Innovation and Universities (Project MAT2015-71664-R and RTI2018-099960-B-I00) and the Serbian Ministry of Ministry of Education, Science and Technological Development (Project No. III45018) for their support. A.P. and Z.K. thank Senzor-INFIZ (Serbia) for the cooperation provided during their respective secondments.